\documentstyle[12pt]{article}
\author{A.V.Berezhnoy\\
{\normalsize \it Nuclear Physics Institute of Moscow State University},
\\{\normalsize \it Moscow, Russia}\\
\vspace*{3mm}
and\\
\vspace*{3mm}
V.V.Kiselev, A.K.Likhoded \\
{\normalsize \it State Research Center of Russia ``Institute of High Energy
Physics''},\\
{\normalsize \it Protvino, Moscow region, 142284, Russia}}
\date{ }
\title{\large\bf New Insight into the Photoproduction of $D^*$-meson in QCD}

\begin{document}
\maketitle
\begin{abstract}
In the framework of the model motivated by perturbative calculations in
the fourth $O(\alpha^3_s\alpha)$-order, the estimates for the
$\gamma p$ cross-section of the $D^*$-meson production in the ZEUS
experiment are performed. We factorize the hadronization of $(c\bar q)$-state
hardly produced in perturbative QCD, which allows us to take into account the
higher twist terms in the powers of $1/p_T$ at $p_T\sim m_c$ and to correctly
reproduce the $c$-quark fragmentation dominant at high $p_T\gg m_c$ to the
given order of $\alpha_s$. We find a good agreement with the experimental data
on the photoproduction, if the color-octet $(c\bar q)$-state is taken into
account, which yields $\langle O_{(8)} \rangle \approx 0.33-0.49 \; {\rm
GeV}^3$.
\end{abstract}

The measurement of the charm production by the ZEUS and H1 collaborations at
HERA in the interaction of quazireal photon with the proton \cite{ZEUS,H1}
rises the problem on the data interpretation in the framework of QCD.
The description of process for the heavy quark production in the
perturbative QCD (pQCD) is completely justified due to the hardness of partonic
subprocesses. At present, there are two types of calculations based on the
results in the next-to-leading order over $\alpha_s$, i.e. in $O(\alpha
\alpha_s^2)$: for the production of massive and massless quarks,
correspondingly \cite{mc,mc0}.

In the both approaches the fragmentation model for the $c$-quark
hadronization into $D^*$ was used at $p_T \gg m_c$. It was applied in the
following form:
\begin{equation}
\frac{d^2\sigma_{D^*}}{dzdp_T}= 
\left. \frac{d\hat \sigma_{c \bar c}(k_T,\mu)}{dk_T}
\right|_{k_T=\frac{p_T}{z}}\cdot \frac{D_{c \to D^*}(z,\mu)}{z},
\end{equation}
where $D_{c \to D^*}(z,\mu)$ is a fragmentation function normalized by
a probability of the $c$-quark fragmentation into the $D^*$-meson, $W(c \to
D^*)$, which is empirically determined from the $e^+e^-$-annihilation
data \cite{LEP}, where $W(c \to D^*)=0.22\pm 0.014\pm0.14$, 
and $\mu \sim p_T$ is the
factorization scale for the perturbative partonic cross section $d\hat
\sigma/dk_{T}$. It is clear that such the approach does not allow one to
describe the experimental data in the region of $p_T \sim m_c$, where the
fragmentation is not in the saturation regime, yet. The fact is valid even for
the case of calculations with the massive quarks, wherein the pQCD can be
correctly applied in the region of $p_T\sim m_c$.

The most important assumption of this method is that the factorization takes
place for the probability of {\sf isolated $c$-quark} hadronization and its
hard production. This circumstance automatically leads to the mentioned
restriction: $p_T \gg m_c$.

In this paper, we discuss a model of the charmed meson production with the
hadronization of $(c\bar q)$-system into the $D^*$-meson instead of the
hadronization of isolated $c$-quark. So, it is quite reasonable to apply this
approach in the region of $p_T\sim m_c$, too, wherein the pQCD can be used.
In the production of $(c\bar q)$-state, it is necessary to take into
account the production of the additional quark pair, so that, in the leading
order of $\alpha_s$, the $O(\alpha\alpha_s^3)$ calculation must be
used. Among the full set of the tree level diagrams on Fig. 1, we can see
several diagrams (i.e. 16 and 19) of heavy quark fragmentation, which dominates
at $p_T\gg m_c$. At high transverse momenta, they reproduce the results of
fragmentation model with the following fragmentation function \cite{frag}:
\begin{eqnarray}
\displaystyle
&&
D_{\bar c \to D^*}(z)=
\frac{8\alpha^2\langle O^{eff} \rangle}{27 m_c^3}
\frac{rz(1-z)^2}{(1-(1-r)z)^6}
(2-2(3-2r)z+3(3-2r+4r^2)z^2
\nonumber
\\
&& \quad
-2(1-r)(4-r+2r^2)z^3+
(1-r)^2(3-2r+2r^2)z^4),
\end{eqnarray}
where $r=m_q/(m_q+m_c)$ and
$$\langle O^{eff} \rangle \approx \langle O_{(1)} \rangle = 
\frac{1}{12 M} \left(-g^{\mu\nu}+\frac {p^\mu p^\nu}{M^2}\right)\;
\langle D^*(p)|(\bar c \gamma_\mu q) (\bar q \gamma_\nu
c)|D^*(p)\rangle,
$$
which corresponds to the squared value of the quarkonium wave function
at the origin for two quarks in the framework of nonrelativistic
potential model: $\langle O_{(1)} \rangle|_{NR} =|\Psi(0)|^2$.
We have introduced the effective mass of light quark $m_q$,
which determines both the normalization and the form of fragmentation
function. It is worth to stress that, in the higher orders of QCD, the
hard part of processes does not depend on $m_q$. However, the problem of
infrared divergences always presents in these cases. Usually, it can be removed
by a redefinition of initial and final states. For the initial states one takes
into account the QCD-evolution of intrinsic quark-gluon sea.
One can introduce a regularization parameter ($\mu_T$ in the NLO-calculations
in \cite{mc,mc0}) to avoid the difficulties connected to the nonperturbative
hadronic sea in the final state. In the model under consideration, $m_q \sim
m_{\rho}/2$ is used as an infrared regulator. In this case the gluonic
virtualities are greater than $|k_g^2|\sim p_T^2+m_{\rho}^2$. So, the fact
allows us to believe in the reliability of pQCD calculation. The $D^*$-meson
production in $e^+e^-$-annihilation has been used to fix the model parameters.
We find that the form and normalization of fragmentation function with the
account for the QCD-evolution is in rather a good agreement with the LEP data,
if the following values of parameters are chosen as below:
\begin{equation}
\begin{array}{rcl}
\alpha_s &=& 0.35,\\
m_q &=& 0.3 \ {\rm GeV},\\
m_c &=& 1.5 \ {\rm GeV},\\
\langle O^{eff} \rangle &=& 0.25 \ {\rm GeV}^{3},\\
W(c\to D^*) &=& 0.22.\\
 \end{array}
\end{equation}

In $e^+e^-$-annihilation the enforcement of color-singlet $(c\bar q)$-state
takes place in comparison with the octet contribution due to the color
amplitude structure, so that
\begin{equation}
\langle O^{eff}\rangle=\langle O_{(1)}\rangle + \frac{1}{8}
\langle O_{(8)} \rangle ,
\end{equation}
where
$$
\langle O_{(8)} \rangle = 
\frac{1}{8 M} \left(-g^{\mu\nu}+\frac{p^\mu p^\nu}{M^2}\right)\;
\langle D^*(p)|(\bar c \gamma_\mu \lambda^a q) (\bar q \gamma_\nu \lambda^b
c)|D^*(p)\rangle\; \frac{\delta^{ab}}{8}.
$$

It is qualitatively clear that the relative contribution of the octet four
quark operator $\langle O_{(8)} \rangle / \langle O_{(1)} \rangle $ is
determined by the relative velocity $v$ of the quark motion in the hadron. For
example, in the hard production of heavy quarkonium the octet fraction is
strongly suppressed due to the nonrelativistic expansion in $v\ll 1$
\cite{NRQCD}. For the meson, containing the light relativistic quark, we can
expect that $\langle O_{(8)} \rangle / \langle O_{(1)} \rangle \sim 1$, since
$v\sim 1$. Supposing this arrangement in $e^+e^-$-annihilation, we see that
$\langle O_{(1)} \rangle$ is dominant in $\langle O^{eff} \rangle$. The
normalization of octet operator $\langle O_{(8)} \rangle$ remains the free
parameter. It can be determined from a process with
another color structure of the production amplitude, for example, from the
photoproduction. In this case, the essential contribution due to diagrams of
nonfragmentation type (recombination) takes place at $p_T \sim
m_c$. We have to emphasize that, in this kinematical region, the octet
contribution is not suppressed in comparison with the singlet one.
  
The calculation results are shown in Fig.~2 in comparison with the experimental
data of ZEUS \cite{ZEUS} in the following kinematical region:
$p_T>2$ GeV, $-1.5<\eta<1.5$, $130\ {\rm GeV}<W<280\ {\rm GeV}$,
$Q^2<1\ {\rm GeV^2}$, where $W$ is the total energy of $\gamma p$ interaction,
$Q^2$ is the photonic virtuality, and $\eta$ is the pseudorapidity of
$D^*$-meson.

We see that the value of the $D^*$-meson production cross section cannot be
explained by the singlet mechanism only. The relative value of octet
contribution $\langle O_{(8)} \rangle / \langle O_{(1)} \rangle = 1.3$ allows
us to describe the spectra of $D^*$-mesons. We would like to emphasize that the
singlet term, say, at $p_T> 4$ GeV is very close to the predictions in the
model for the production of isolated $c$-quark \cite{mc,mc0}.
 
Following a request by the ZEUS Collaboration we have calculated the
differential cross sections for the $D^*$-meson production, as predicted by the
model under consideration and shown in Fig.~3,
at low $W$ for the following kinematical region: $p_T>2$ GeV, $-1.<\eta<1.5$,
$80\ {\rm GeV}<W<120{\rm GeV}$, $Q^2<0.01\ {\rm GeV^2}$. The same values of
$\langle O_{(1)} \rangle$ and $\langle O_{(8)} \rangle$ have been used in the
both calculations. 

Using the experimental data by the ZEUS collaboration, only, we cannot exclude
the possibility of other ratio between the singlet and octet operators, if the
octet fraction is not suppressed in the fragmentation regime in
$e^+e^-$-annihilation. Then, for example, at $\langle O_{(1)} \rangle \simeq
0.14\ {\rm GeV}^3$ and $\langle O_{(8)} \rangle / \langle O_{(1)} \rangle \sim
3.5$ we find a good description of the experimental data (see Fig.~4). However,
the low value of singlet operator in combination with the high fraction of
octet contribution results in quite a slow variation of the octet term.

Further, in the calculations we have neglected the differences between the
spectra of color-octet vector and pseudoscalar states, whose contributions have
been effectively summed up in the corresponding operator. Moreover, we have not
considered a possible influence by the octet $P$-wave states, which seem to be
very unstable for the heavy-light system.

In conclusion, we have presented the model for the charmed meson production in 
photon-proton collisions with the hadronization of $(c\bar q)$-state in
contrast to the usual restriction by the isolated $c$-quark. It allows us to
make justified predictions in the range of $p_T\sim m_c$, which improves the
results of simple $c$-quark fragmentation regime dominant at high transverse
momenta. In addition, the model includes the contributions of color-singlet and
octet $(c\bar q)$-states, which depend on the color structure of production
amplitude. So, in $e^+e^-$-annihilation the octet fraction is suppressed, while
in the photoproduction we have found that the singlet contribution
underestimates the data, and the octet operator can be measured so that, in the
conservative evaluation, we extract
$$
\langle O_{(8)} \rangle \approx 0.33-0.49 \; {\rm GeV}^3.
$$
To improve the estimation, we plan to analyze the model predictions for the
electroproduction, where the photon can be essentially virtual, in comparison
with the current data.

This work is in part supported by the Russian Foundation for Basic Research,
grants 96-02-18216 and 96-15-96575. 

The authors would like to express the gratitude to Prof. A.Wagner and
members of DESY Theory Group for their kind hospitality during the visit 
to DESY, where this work was elaborated, as well as to Prof. S.S.Gershtein 
for discussions. The authors thank Prof. G.Kramer for conversation and remarks,
which essentially improved the presentation of paper.

We thank Leonid Gladilin (ZEUS) for useful discussions, explanations on the
data sample and help in our research.

\newpage
\includegraphics{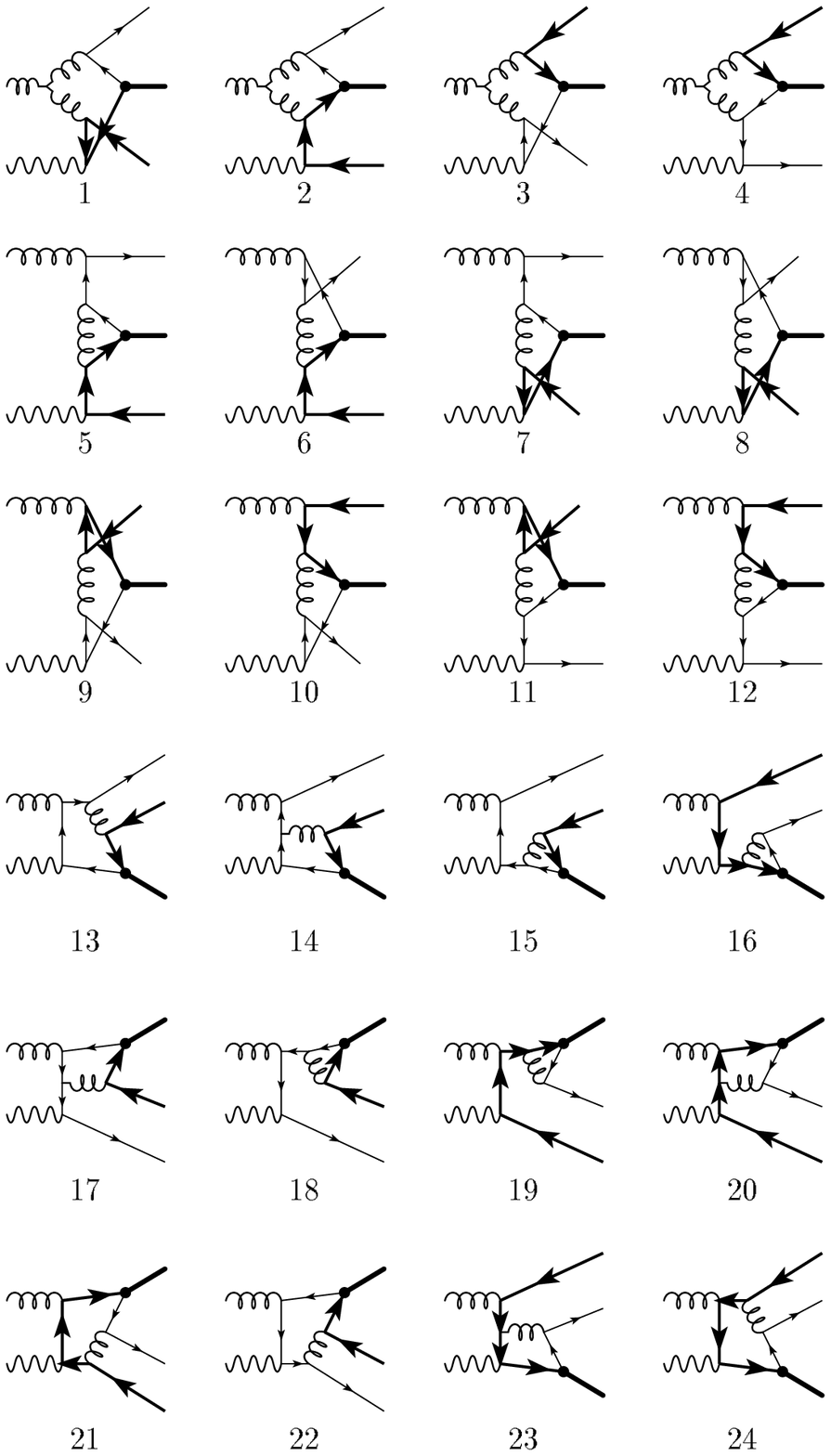}
\vspace*{21 cm}
\hspace*{3 cm} Fig.~1. \parbox[t]{8cm}{The leading order diagrams for the
$(c\bar q)$-state production in $g \gamma$ interactions.}

\newpage
\vspace*{-3 cm}
\includegraphics{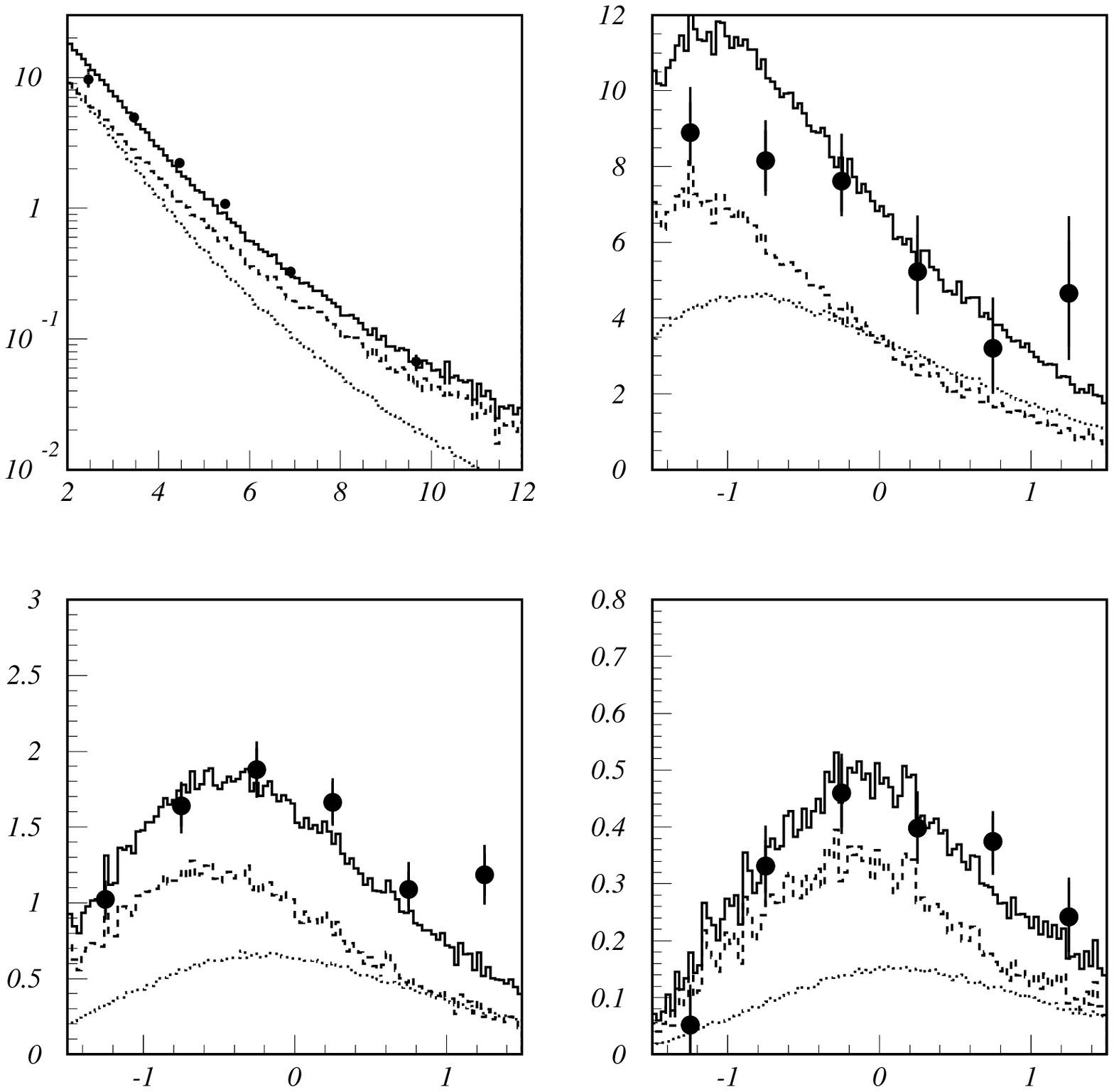}

\begin{picture}(450,450)
\put(-10,350){$d\sigma/dp_T$, nb/GeV}
\put(240,350){$d\sigma/d\eta$, nb}
\put(-10,100){$d\sigma/d\eta$, nb}
\put(240,100){$d\sigma/d\eta$, nb}
\put(150,120){$p_T$, GeV}
\put(400,120){$\eta$}
\put(150,-130){$\eta$}
\put(400,-130){$\eta$}
\put(345,300){$p_T>2$ GeV}
\put(345,50){$p_T>6$ GeV}
\put(95,50){$p_T>4$ GeV}
\put(20,-180){Fig.~2. \parbox[t]{10cm}{The distributions of $D^{*\pm}$-meson
photoproduction for $\langle O_8 \rangle / \langle O_1 \rangle =1.3$
in comparison with the ZEUS data at $130\ {\rm GeV}<W<280 \ {\rm GeV}$ and
$Q^2<1\ {\rm GeV^2}$. The color-singlet term is shown by the dashed histogram,
the octet contribution is given by the dotted histogram, and the solid
histogram presents the sum of two terms.}}
\end{picture}

\newpage
\vspace*{-3cm}
\includegraphics{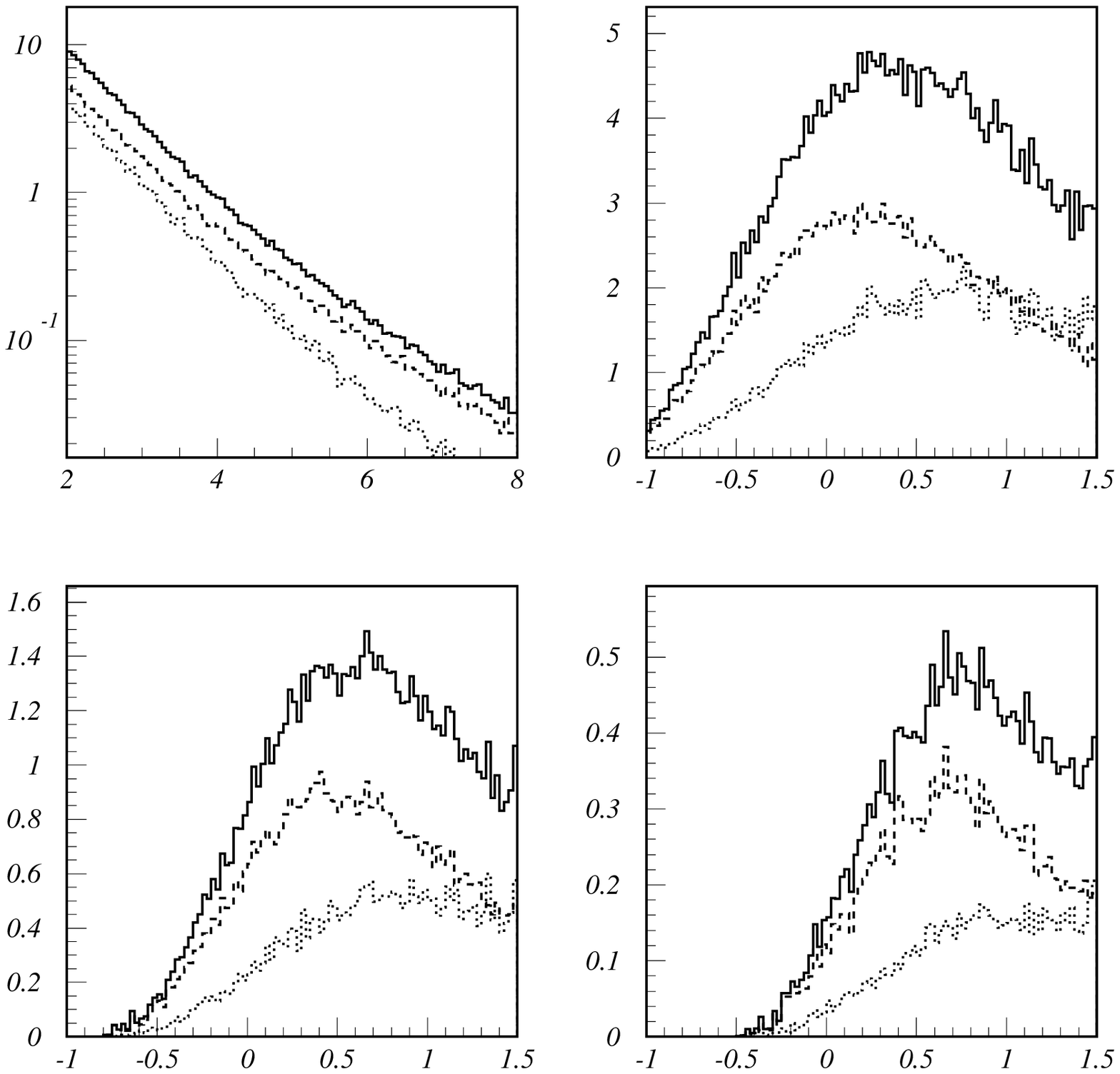}

\begin{picture}(450,450)
\put(-10,350){$d\sigma/dp_T$, nb/GeV}
\put(240,350){$d\sigma/d\eta$, nb}
\put(-10,100){$d\sigma/d\eta$, nb}
\put(240,100){$d\sigma/d\eta$, nb}
\put(150,120){$p_T$, GeV}
\put(400,120){$\eta$}
\put(150,-130){$\eta$}
\put(400,-130){$\eta$}
\put(245,310){$p_T>2$ GeV}
\put(0,60){$p_T>3.25$ GeV}
\put(245,60){$p_T>4.5$ GeV}
\put(20,-180){Fig.~3. \parbox[t]{10cm}{The distributions of $D^*$-meson
photoproduction for $\langle O_8 \rangle / \langle O_1 \rangle =1.3$ at
$80\ {\rm GeV}<W<120 \ {\rm GeV}$ and $Q^2<0.01\ {\rm GeV^2}$. The notations
are the same as in Fig.~2.}}
\end{picture}

\newpage
\vspace*{-3 cm}
\includegraphics{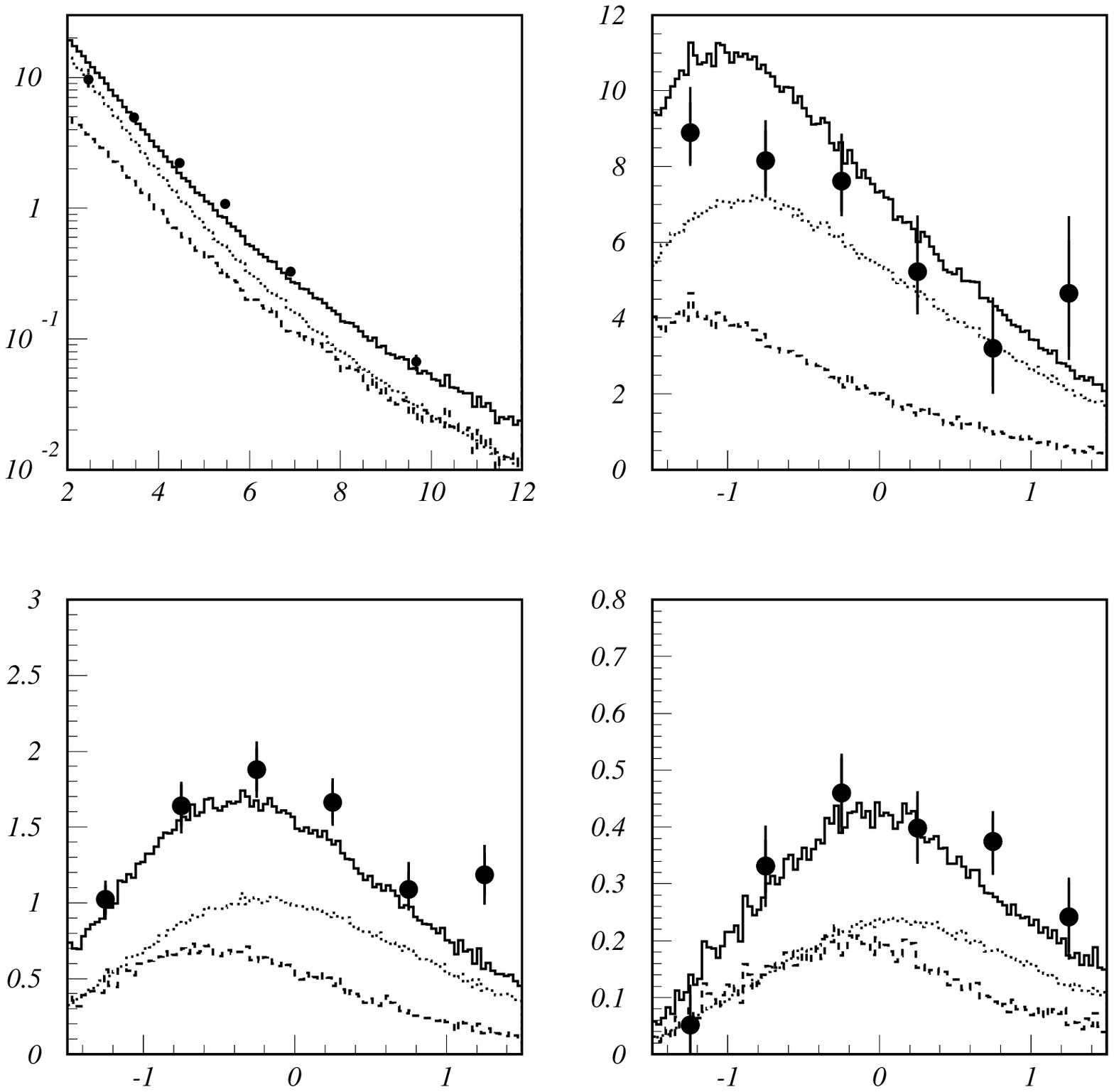}

\begin{picture}(450,450)
\put(-10,350){$d\sigma/dp_T$, nb/GeV}
\put(240,350){$d\sigma/d\eta$, nb}
\put(-10,100){$d\sigma/d\eta$, nb}
\put(240,100){$d\sigma/d\eta$, nb}
\put(150,120){$p_T$, GeV}
\put(400,120){$\eta$}
\put(150,-130){$\eta$}
\put(400,-130){$\eta$}
\put(345,300){$p_T>2$ GeV}
\put(345,50){$p_T>6$ GeV}
\put(95,50){$p_T>4$ GeV}
\put(20,-180){Fig.~4. \parbox[t]{10cm}{The distributions of $D^*$-meson
photoproduction for $\langle O_8 \rangle / \langle O_1 \rangle =3.5$
in comparison with the ZEUS data at $130\ {\rm GeV}<W<280 \ {\rm GeV}$ and
$Q^2<1\ {\rm GeV^2}$. The notations are the same as in Fig.~2.}}
\end{picture}

\newpage
\vspace*{-3cm}
\includegraphics{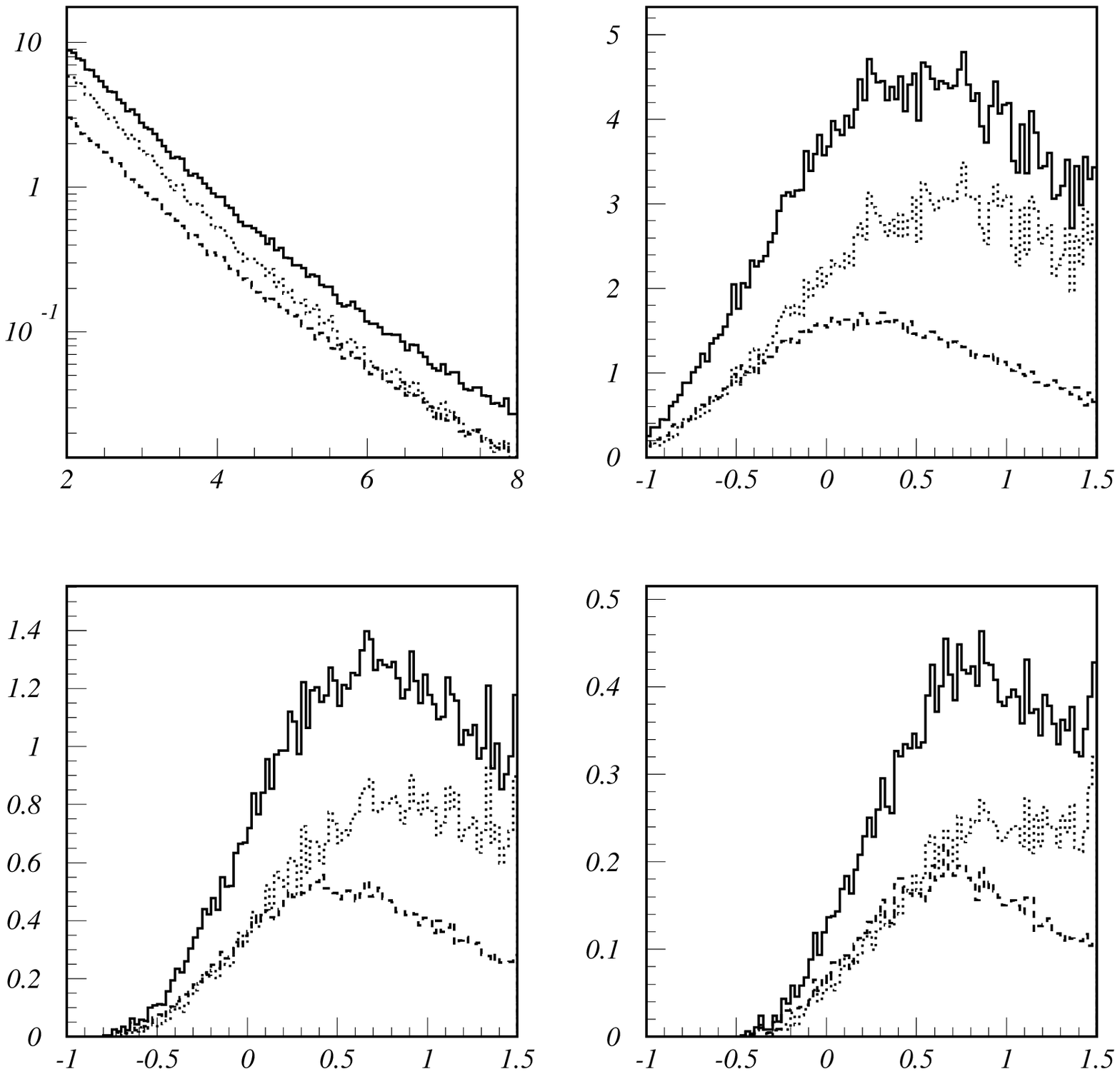}

\begin{picture}(450,450)
\put(-10,350){$d\sigma/dp_T$, nb/GeV}
\put(240,350){$d\sigma/d\eta$, nb}
\put(-10,100){$d\sigma/d\eta$, nb}
\put(240,100){$d\sigma/d\eta$, nb}
\put(150,120){$p_T$, GeV}
\put(400,120){$\eta$}
\put(150,-130){$\eta$}
\put(400,-130){$\eta$}
\put(245,310){$p_T>2$ GeV}
\put(0,60){$p_T>3.25$ GeV}
\put(245,60){$p_T>4.5$ GeV}
\put(20,-180){Fig.~5. \parbox[t]{10cm}{The distributions of $D^*$-meson
photoproduction for $\langle O_8 \rangle / \langle O_1 \rangle =3.5$ at
$80\ {\rm GeV}<W<120 \ {\rm GeV}$ and $Q^2<0.01\ {\rm GeV^2}$. The notations
are the same as in Fig.~2.}}
\end{picture}

\end{document}